\documentclass[aps,prl,twocolumn,groupaddress,showpacs,floatfix]{revtex4}

\usepackage{txfonts}
\usepackage{graphicx}

\begin{document}

\title{Quantum Phase Diffusion in a Small Underdamped Josephson Junction}
\author{H. F. Yu, X. B. Zhu, Z. H. Peng, Ye Tian, D. J. Cui, G. H. Chen, D.
N. Zheng, X. N. Jing, Li Lu, and S. P. Zhao}
\affiliation{Beijing National Laboratory for Condensed Matter Physics, Institute of
Physics, Chinese Academy of Sciences, Beijing 100190, China}
\author{Siyuan Han}
\affiliation{Department of Physics and Astronomy, University of Kansas, Lawrence, Kansas
66045, USA}

\begin{abstract}
Quantum phase diffusion in a small underdamped Nb/AlO$_x$/Nb junction ($\sim$
0.4 $\mu$m$^2$) is demonstrated in a wide temperature range of 25-140 mK
where macroscopic quantum tunneling (MQT) is the dominant escape mechanism.
We propose a two-step transition model to describe the switching process in
which the escape rate out of the potential well and the transition rate from
phase diffusion to the running state are considered. The transition rate
extracted from the experimental switching current distribution follows the
predicted Arrhenius law in the thermal regime but is greatly enhanced when
MQT becomes dominant.
\end{abstract}

\maketitle





Classical and quantum diffusion of Brownian particles in titled periodic
potential plays a fundamental role in the dynamical behavior of many systems
in science and engineering \cite%
{mart89,ians89,vion96,kova04,mann05,kivi05,kras05,li07,fent08,evs08,lee06,gom08,sch03,tay10,hag09,kot05}%
. Examples include current biased Josephson junctions \cite%
{mart89,ians89,vion96,kova04,mann05,kivi05,kras05,li07,fent08},
colloidal particles in arrays of laser traps \cite{evs08,lee06},
cold atoms in optical lattice or Bose-Einstein condensates
\cite{gom08,sch03,tay10}, and various biology-inspired systems known
as Brownian motors (molecular motors or life engines), which receive
considerable attention in physics \cite{hag09} and chemistry
\cite{kot05}. Because of the design flexibility, manufacturability,
and controllability Josephson junctions provide an excellent test
bed for making quantitative comparison of experimental data with
theoretical predictions and unraveling possible new physics in the
tilted periodic potential systems.

The dynamics of a current biased Josephson junction can be
visualized as a
fictitious phase particle of mass $C$ moving in a tilted periodic potential $%
U(\varphi )=-E_{J}(i\varphi +\cos \varphi )$. Here, $C$ is junction
capacitance, $i=I/I_{c}$ is the junction's bias current normalized to its
critical current, the phase particle's position $\varphi $ is the gauge
invariant phase difference across the junction, and $E_{J}=\hbar I_{c}/2e$
is the Josephson coupling energy with $e$ and $\hbar $ being the electron
charge and Planck's constant, respectively. Previous experiments using
Josephson junctions have identified three distinctive dynamical states, as
shown schematically in \textrm{Fig.~1}. In the first state, the phase
particle is trapped in one of the metastable potential wells and undergoes
small oscillation around the bottom of the well with plasma frequency $%
\omega _{p}$. Because of thermal and/or quantum fluctuations the
particle has a finite rate $\Gamma _{1}$ escaping from the trapped
state. The escape rate becomes significant when the barrier height
$\Delta U$ is not much greater than $k_{B}T$ or $\hbar \omega _{p}$,
where $k_{B}$ is the Boltzmann constant and $T$ denotes the
temperature, respectively. After the particle escapes from the
initial well, depending on the energy gain $\delta U=\Phi _{0}I$
($\Phi _{0}$ being the flux quantum) and the loss $E_{D}$ due to
damping (cf. \textrm{Fig. 1}), it could enter either the second
dynamical state called phase diffusion (PD) or the final running
state. In the former case as the bias current $I$ is increased
further the particle will eventually make a transition,
characterized by a rate constant $\Gamma _{2},$ to the running
state. While escape from the trapped state to PD is difficult to
detect transition to the running state is signaled by a sudden jump
in the dc voltage of the junction (called switching) and thus can be
readily captured in real time by increasing $I$ continuously from
zero until a switching occurs \cite{fult74}.

\begin{figure}[b]
\includegraphics[width=0.35\textwidth]{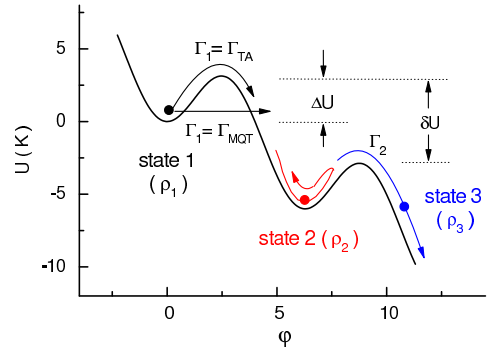}
\caption{(Color online) Phase particle in the trapped, diffusion,
and running states (denoted by $n=1,2,3,$ respectively) with
occupation probability $\protect\rho _{n}$ in a tilted washboard
potential.} \label{Fig-1}
\end{figure}

The fundamental importance of understanding PD has stimulated many
studies in recent years. However, experimental studies were focused
mostly on the classical regime where thermal activation (TA) is the
dominant escape mechanism and thermal fluctuation governs the PD
process \cite
{mart89,ians89,vion96,kova04,mann05,kivi05,kras05,li07,fent08}. On
the other hand, in the quantum regime where macroscopic quantum
tunneling (MQT) dominates, one expects that quantum fluctuation
induced tunneling will play an important role in the PD process and
subsequent transition to the running state thus the term quantum PD
(QPD) has been coined in the literature
\cite{hag09,anke04,mach06,deni09}. However, although theoretical
progress of QPD in overdamped systems has been remarkable over
recent years \cite{anke04,mach06} the situation is so far much less
clear for underdamped systems \cite{hag09,deni09}.

\begin{table}[b]
\caption{Parameters of two Nb/AlO$_{x}$/Nb junctions S and L used in
this work. $R_{N}$ is normal-state resistance obtained from $I$-$V$
curves. $I_{c}$, $C$, and $R$ for L are determined from fits to
experiment using TA and MQT theories below 450 mK and Monte Carlo
simulations above it. Those for S are obtained considering its
$R_{N}$ ratio to L (Note a slightly larger $R$ chosen to have a
better fit). See the text for details.}
\label{Table-1}%
\begin{ruledtabular}
\begin{tabular}{cccccccc}
Junction&Area\footnote{Estimated for L from fitted $C$ and a
specific capacitance of 50 fF/$\mu$m$^2$. The value for S is
obtained via its $R_{N}$ ratio to L. Nominal areas for junctions S
and L were 0.52
and 1.61 $\mu$m$^2$, respectively.}($\mu$m$^2$)&$R_N$(k$\Omega$)&$I_c$(nA)&$C$(fF)&$R$($\Omega$)&$T_{cr}$(mK)&$T_0$(mK)\\
\colrule S &0.39 &15.1 &122 &19.6 &1800 &140 &$<$ 25\\
         L &1.54 &3.84 &480 &77   &315  &125 &$\sim$450
\end{tabular}
\end{ruledtabular}
\end{table}

In this work, we demonstrate QPD in a small underdamped Josephson
junction over a wide temperature range of $25$ to $140$ mK. To
contrast QPD with classical PD, we use two Nb-AlO$_{x}$-Nb trilayer
junctions of different sizes (see Table I) having $T_{0}$ $\ll $
$T_{cr}$ and $T_{0}$ $\gg $ $T_{cr} $, respectively. Here, $T_{0}$
is the temperature above which PD occurs and $T_{cr}$ is the
classical-to-quantum crossover temperature below which MQT
dominates. One of the hallmarks of PD in underdamped junctions is
the narrowing of the width $\sigma$ of switching current
distribution $P(I)$ as temperature increases
\cite{mann05,kivi05,kras05,li07}. This is observed clearly in the
measured $\sigma (T)$ of the larger junction L above
$T_{0}^{L}\simeq 450$ mK $\gg T_{cr}^{L}$, which indicates that PD
in this case is classical in nature. In sharp contrast, for the
smaller junction S the width $\sigma$ continues to increase as
temperature decreases to the lowest value of 25 mK. When plotted in
semilogarithmic scale $\sigma$ vs $T$ shows a clear increase of
slope around $T_{cr}^{S}=140$ mK, pointing to a change from
classical PD to QPD. We will extract the transition rate
$\Gamma_{2}$ directly from the experimental results and show that
QPD is fundamentally different from classical PD.

Two Nb/AlO$_{x}$/Nb junctions used in this study were fabricated on
the same chip with nominal areas of $0.52$ and $1.61$ $\mu $m$^{2}$
for junctions S and L, respectively. Compared with previous works
reported in Refs.~\cite{mann05} and \cite{kivi05}, where dc SQUIDs
were used to tune $I_{c}$, our approach kept $I_{c}/C$ constant.
This unique approach is essential to extend PD to the quantum
regime. Since $T_{cr}=\hbar \omega
_{p}[(1+1/4Q^{2})^{1/2}-1/2Q]/2\pi k_{B}$ $\sim $ $\hbar \omega
_{0}/2\pi k_{B}$ scales with the plasma frequency $\omega
_{p}=\omega _{0}\left( 1-i^{2}\right) ^{1/4},$ where $\omega
_{0}=(2\pi I_{c}/\Phi _{0}C)^{1/2}$ and $Q=\omega _{p}RC$ ($R$ being
junction's damping resistance), $T_{cr}$ is approximately
independent of the junction sizes as long as they are fabricated
from the same trilayer. On the other hand, $T_{0}$ can be reduced by
making smaller junctions therefore we are able to tune $T_{0}$ and
$T_{cr}$ independently to meet the condition $T_{0}\ll T_{cr}$
required for observing QPD \cite{rem}.

\begin{figure}[t]
\centering \includegraphics[width=0.38\textwidth]{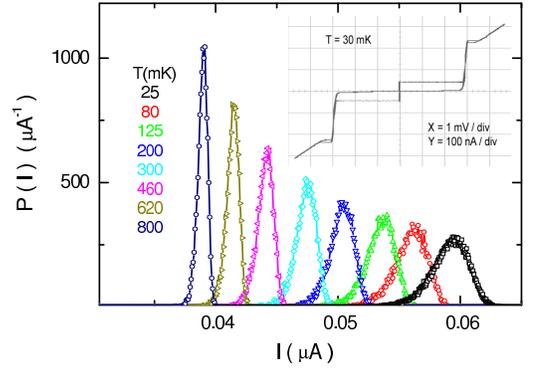}
\caption{(Color online) Experimentally measured $P(I)$ of junction S
at some temperatures indicated. The inset shows the $I$-$V$ trace of
the junction at 30 mK.}
\end{figure}

\begin{figure}[b]
\centering \includegraphics[width=0.35\textwidth]{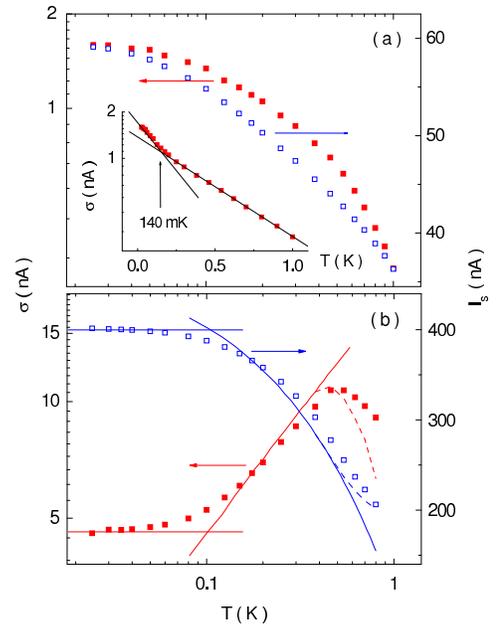}
\caption{(Color online) (a) Width $\protect\sigma$ and mean $I_s$ of
experimental $P(I)$ of junction S (symbols). (b) Corresponding data
of junction L. Solid lines in (b) are calculated from TA and MQT
theories while dashed lines from Monte Carlo simulations considering
thermal PD \protect\cite{mann05,li07}. The inset shows
$\protect\sigma$ of junction S plotted in semilogarithmic scale. Two
solid lines are guides to the eye displaying a slope turning near
$T_{cr}^S$ = 140 mK.}
\end{figure}

Figure~2 shows the measured $P(I)$ from $25$ to $800$ mK for
junction S with its $I$-$V$ curve at $30$ mK displayed in the inset.
In our experiment, $P(I)$ was measured by the time-of-flight
technique \cite{li07,yu10} with $di/dt$ $=110$/sec for sample S and
$163$/sec for sample L. Each measured $P(I)$ consisted of $50000$
switching events. In Fig.~3, we plot $\sigma $ and the mean $I_{s}$
of $P(I)$ versus temperature (symbols) for junction S in (a)
together with those of junction L in (b). For junction L the
measured $\sigma (T)$ shows the familiar classical PD started at
temperature $T_{0}^{L}$ $\simeq$ $450$ mK well above $T_{cr}^{L}$ =
125 mK. The solid lines in (b) are calculated according to the TA
\cite{kram40} and MQT \cite{cald81} rate formulas using the
parameters listed in Table I. The dashed lines are from Monte Carlo
simulations considering thermal fluctuation and PD
\cite{mann05,li07}. In contrast to junction L the observed $\sigma $
for junction S in Fig. 3(a) shows a monotonic decrease with
increasing temperature, indicating that PD occurred in the entire
temperature range of the experiment. Furthermore when plotting the
data in semilogarithmic scale as shown in the inset of Fig.~3 we
notice a distinctive slope decrease around $T_{cr}^{S}$ = 140 mK
from MQT to TA regimes. Such a decrease can be easily understood
since TA causes $\sigma $ to increase with increasing $T$ which
partially cancels the effect of negative $(1/\sigma)d\sigma /dT$ due
to PD.

To gain further insight and have a quantitative grasp on the effects
of escape (from the trapped state to PD) and transition (from PD to
the running state) on switching current distribution, regardless of
whether TA or MQT is the dominant mechanism, we set up the following
master equation according to the two-step transition model shown in
Fig.~1:
\begin{equation}
\left\{
\begin{array}{r@{\quad = \quad}l}
d\rho _{1}/dt= & -\Gamma _{1}~\rho _{1} \\
d\rho _{2}/dt= & \Gamma _{1}~\rho _{1}-\Gamma _{2}~\rho _{2} \\
d\rho _{3}/dt= & \Gamma _{2}~\rho _{2},%
\end{array}%
\right.  \label{MasterEq}
\end{equation}%
where $\rho _{n}$ ($n$ = 1, 2, 3) is the probability of finding the
phase particle in state $n$. Since $P(I)$ $=$ $d\rho _{3}/dI$, it
follows straightforwardly that
\begin{equation}
\Gamma _{2}(I)=\frac{(dI/dt)P(I)}{1-\int_{0}^{I}P(I^{^{\prime
}})dI^{^{\prime }}-e^{-\frac{1}{dI/dt}\int_{0}^{I}\Gamma _{1}(I^{^{\prime
}})dI^{^{\prime }}}}~.  \label{Gamma2}
\end{equation}
Equation (\ref{Gamma2}) shows that $\Gamma _{2}(I)$ can be extracted
from measured $P(I)$ provided $\Gamma _{1}(I)$ is known, which is
true in our experiment. Notice that in the limit of $\Gamma _{2}$
$\rightarrow $ $\infty $, Eq.~(\ref{Gamma2}) leads directly to
$\Gamma _{1}(I)=(dI/dt)P(I)/[1-\int_{0}^{I}P(I^{^{\prime
}})dI^{^{\prime }}]$ which is identical to the result of Fulton and
Dunkleberger \cite{fult74} in which PD is absent. In the opposite
limit of $\Gamma _{2}$ $\ll $ $\Gamma _{1}$, the same expression is
obtained with $\Gamma _{1}$ replaced by $\Gamma _{2}$: $\Gamma
_{2}(I)=(dI/dt)P(I)/[1-\int_{0}^{I}P(I^{^{\prime }})dI^{^{\prime
}}]$. These results mean that the much slower process plays the
major role in determining $P(I)$, as expected. In the more general
situation of $\Gamma _{2}\sim \Gamma _{1}$, Eq.~(2) enables one to
separate the effect of $\Gamma _{2}$ on switching current
distributions from that of $\Gamma _{1}$. The inverse procedure of
computing $P(I)$ from $\Gamma _{1}$ and $\Gamma _{2}$ is given by:
\begin{equation}
P(I)=\frac{\Gamma _{2}}{(dI/dt)^{2}}e^{-\frac{1}{dI/dt}\int_{0}^{I}\Gamma
_{2}dI^{^{\prime }}}\int_{0}^{I}\Gamma _{1}e^{-\frac{1}{dI/dt}%
\int_{0}^{I^{^{\prime }}}(\Gamma _{1}-\Gamma _{2})dI^{^{\prime \prime
}}}dI^{^{\prime }}.  \label{P(I)}
\end{equation}
Equations (\ref{Gamma2}) and (\ref{P(I)}) thus allow us to
quantitatively investigate the dependence of (Q)PD on bias current
and the interplay between the particle's escape and (Q)PD. In
Fig.~4(a), we plot $\Gamma _{1}$ (solid lines) calculated using the
parameters of junction S and $\Gamma _{2}$ (symbols) extracted from
the measured $P(I)$ using Eq.~(\ref{Gamma2}). It can be seen that at
$T$ $=800$ mK, $\Gamma _{1}$ is several orders of magnitude greater
than $\Gamma _{2}$. The measured $P(I)$ is therefore entirely
determined by $\Gamma _{2}$. As the temperature decreases, $\Gamma
_{1}$ is seen to progressively approach $\Gamma _{2}$.

Having clearly established that PD occurs in both classical and
quantum regimes in junction S, we now use the data in Fig.~4(a) to
further demonstrate the key difference between classical PD and QPD.
In Fig.~4(b), we plot $\Gamma _{2}$ versus $1/T$ at three bias
currents (thus fixed potentials) of 48, 52, and 56 nA, which shows
distinct features below and above $T_{cr}^{S}.$ While the data above
$T_{cr}^{S}$ follow the straight lines, indicating that $\Gamma
_{2}$ in the classical regime obeys the Arrhenius law $\Gamma _{2}$
displays a much weaker $1/T$ dependence at below $T\ll T_{cr}^{S}$.
We note that similar behavior in the classical regime was discussed
previously by Vion \textit{et al.} \cite{vion96} for overdamped
system where the diffusive particle is considered to overcome an
effective dissipation barrier. In that case, the transition rate
from PD to the running state, which retains the familiar Kramers
form, was derived. Fitting the data above $T_{cr}^{S}$ using $\Gamma
_{2}$ $=$ $a\exp (-b/T)$, we obtain $a$ $=5.2\times 10^{7}$
sec$^{-1}$, $b$ $=2.3$ K for $I$ $=48$ nA (dashed line) and $a$
$=3.3\times 10^{8}$ sec$^{-1}$, $b$ $=1.7$ K for $I$ $=52$ nA
(dotted line). The effective barrier $b$ appears smaller as compared
to the calculated barrier height $\Delta U$ of $2.68$ and $2.46$ K
due to the motion of the diffusive particles, which is physically
quite reasonable. These results indicate that in the thermal regime
a dissipation-barrier description is also applicable to PD in
underdamped junctions.

\begin{figure}[t]
\centering \includegraphics[width=0.32\textwidth]{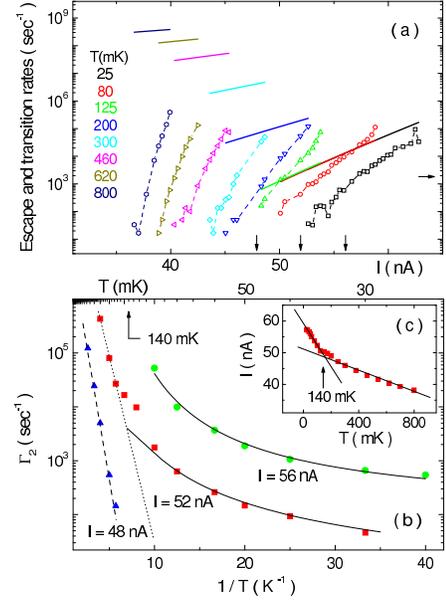}
\caption{(Color online) (a) Transition rate $\Gamma _{2}$ (symbols)
and escape rate $\Gamma _{1}$ (solid lines) of junction S at some
typical temperatures. (b) $\Gamma _{2}$ $\sim$ $1/T$ at three fixed
currents as indicated by the vertical arrows in (a). Dashed and
dotted lines are fits displaying the Arrhenius law. (c) $I$ $\sim$
$T$ for fixed $\Gamma _{2}$ = 2000 sec$^{-1}$ as indicated by a
horizontal arrow in (a). Solid lines in (b) and (c) are guides to
the eye.}
\end{figure}

Machura \textit{et al.} recently investigated the diffusion problem
of overdamped particles using the Smoluchowski equation
incorporating quantum fluctuations \cite{mach06}. They found that
the particle's average velocity $\langle v\rangle $ increases with
increasing temperature and quantum effects always assist the
particle to overcome barriers leading to a larger $\langle v\rangle
$ than that in absence of quantum fluctuations. Because in our
underdamped junction the dc voltage, which is proportional to
$\langle v\rangle ,$ produced by PD is too low to be detected
directly \cite{rem1}, it can nevertheless be expected that a larger
$\langle v\rangle $ would result in a larger $\Gamma _{2}$ since the
increased kinetic energy makes transitions to the running state
easier. For this reason, the data in Fig.~4(b) are consistent with
the theoretical prediction since extrapolating $\Gamma _{2}$ from
the classical to the quantum regime would lead to rates that are
much lower than the experimental data. Therefore, the much weaker
$1/T$ dependence of $\Gamma _{2}$ below $T_{cr}^{S}$, in a stark
contrast to the Arrhenius behavior above $T_{cr}^{S}$, manifests the
quantum nature of the diffusion process at $T$ $<$ $T_{cr}^{S}$.

In Fig. 4(c) we plot $I$ versus $T$ for a constant $\Gamma_{2}=2000$
sec$^{-1}$, which again shows a distinctive change of slope around
$T_{cr}^{S}$ similar to that of $\sigma $. The approximate linear
$I$ $-T$ dependence above $T_{cr}^{S}$ can be qualitatively
explained. In the absence of thermal fluctuations transition from PD
to running state is expected to occur deterministically at $I_{0}$
where $\delta U_{0}=(h/2e)I_{0}=E_{D}.$ For $T>0$ the phase particle
will exit the PD state prematurely because the particle on average
acquires an additional thermal energy of $\sim$ $k_{B}T$. Thus the
condition for transition out of PD needs to be revised to $\delta
U+k_{B}T=E_{D}.$ Assuming the junction's damping, and thus $E_{D},$
saturates at low $T$ we obtain $(h/2e)I=E_{D}-k_{B}T.$ The predicted
slope $|s|=2ek_{B}/h\approx 7$ nA/K is comparable to the
experimental value of $15$ nA/K in the thermal regime in Fig.~4(c),
which is quite reasonable considering the simplicity of the model.
Below $T_{cr}^{S}$, however, the measured $|s|$ increased to about
$68$ nA/K, about an order of magnitude greater than $2ek_{B}/h$
which remains unexplained.

In conclusion, QPD was demonstrated and systematically studied in a
small underdamped Nb Josephson junction. Using junctions of
different sizes fabricated on the same chip we were able to
calibrate the relevant parameters of the small junction and at the
same time extended QPD over a wide temperature range. We showed that
$\sigma $ decreases monotonically with increasing temperature and
there is a distinctive change of slope at $T_{cr}$ below and above
which QPD and classical PD occur. We developed a two-step transition
model with which the effects of escape rate $\Gamma _{1}$ (from the
trapped state) and the transition rate $\Gamma _{2}$ (from PD to the
running state) on switching current distributions can be separated
and $\Gamma _{2}$ be determined from the measured $P(I)$ directly.
It was found that $\Gamma _{2}$ vs $T$ at fixed bias current, and
thus fixed potential landscape, follows the Arrhenius law in the
case of classical PD. The most important finding was that for QPD,
$\Gamma _{2}$ is exponentially higher than that expected for the
classical PD and has a much weaker 1/$T$ dependence. The
similarities between the temperature dependence of $\Gamma _{1}$ and
$\Gamma _{2}$ in underdamped Josephson junctions going from
classical regime to quantum regime were striking. We hope our
experimental progress and advancement in data analysis will
stimulate further theoretical and experimental studies of and lead
to a better understanding of the quantum diffusion phenomena in
underdamped tilted periodic potential systems.

We are grateful to H. Tang and Z. B. Su for helpful discussion and
derivation of Eqs.~(2) and (3). We thank V. Patel, W. Chen, and J.
E. Lukens for providing us with the samples used in this work. The
work at the Institute of Physics was supported by NSFC (Grant No.
10874231) and 973 Program (Grants No. 2009CB929102 and No.
2011CBA00106). S. Han was supported in part by NSF Grant No.
DMR-0325551.



\begin{thebibliography}{99}
\bibitem{rem} In the case of dc SQUIDs we have fixed $C$ thus $T_{cr}$ $%
\propto $ $\sqrt{E_{J}}$ which decreases also as $I_{c}$ is reduced.
\bibitem{rem1} No dc voltage signal within our measurement sensitivity
($\sim $ 0.5 $\mu $V) was detected for junction S before switching.
\end{thebibliography}
\end{document}